\documentclass[letterpaper,twocolumn,english,aps,prl,superscriptaddress]{revtex4-1}
\usepackage{ae,aecompl}
\usepackage[T1]{fontenc}
\usepackage{textcomp}
\usepackage{amstext}
\usepackage{graphicx}
\usepackage[dvipdfm]{hyperref}
\hypersetup{
   colorlinks,
   linkcolor=blue,
   citecolor=blue,
   urlcolor=blue,
}

\makeatletter

\newcommand{\lyxmathsym}[1]{\ifmmode\begingroup\def\b@ld{bold}
  \text{\ifx\math@version\b@ld\bfseries\fi#1}\endgroup\else#1\fi}

\@ifundefined{textcolor}{}
{%
 \definecolor{BLACK}{gray}{0}
 \definecolor{WHITE}{gray}{1}
 \definecolor{RED}{rgb}{1,0,0}
 \definecolor{GREEN}{rgb}{0,1,0}
 \definecolor{BLUE}{rgb}{0,0,1}
 \definecolor{CYAN}{cmyk}{1,0,0,0}
 \definecolor{MAGENTA}{cmyk}{0,1,0,0}
 \definecolor{YELLOW}{cmyk}{0,0,1,0}
 }

\makeatother

\usepackage{babel}
\begin{document}

\title{Fragility of ferromagnetic double exchange interactions and pressure tuning of magnetism in 3d-5d double perovskite  $\mathrm{Sr_{2}FeOsO_{6}}$}

\author{L. S. I. Veiga}

\affiliation{Advanced Photon Source, Argonne National Laboratory, Argonne, Illinois
60439, USA}

\affiliation{Laboratório Nacional de Luz Síncrotron - P.O. Box 6192, 13084-971,
Campinas, São Paulo, Brazil}

\affiliation{Instituto de Física {}``Gleb Wataghin'', Universidade Estadual
de Campinas, Campinas, São Paulo 13083-859, Brazil}

\author{G. Fabbris}

\affiliation{Advanced Photon Source, Argonne National Laboratory, Argonne, Illinois
60439, USA}

\affiliation{Department of Physics, Washington University, St. Louis, Missouri
63130, USA}

\author{M. van Veenendaal}

\affiliation{Advanced Photon Source, Argonne National Laboratory, Argonne, Illinois
60439, USA}

\affiliation{Department of Physics, Northern Illinois University, De Kalb, Illinois
60115, USA}

\author{N. M. Souza-Neto}

\affiliation{Laboratório Nacional de Luz Síncrotron - P.O. Box 6192, 13084-971,
Campinas, São Paulo, Brazil}

\author{H. L. Feng}

\affiliation{Superconducting Properties Unit, National Institute for Materials
Science, 1-1 Namiki, Tsukuba, Ibaraki, 305-0044, Japan}

\affiliation{Graduate School of Chemical Science and Engineering, Hokkaido University,
Sapporo, Hokkaido, 060-0810, Japan}

\author{K. Yamaura}

\affiliation{Superconducting Properties Unit, National Institute for Materials
Science, 1-1 Namiki, Tsukuba, Ibaraki, 305-0044, Japan}

\affiliation{Graduate School of Chemical Science and Engineering, Hokkaido University,
Sapporo, Hokkaido, 060-0810, Japan}

\author{D. Haskel}

\email{haskel@aps.anl.gov}

\affiliation{Advanced Photon Source, Argonne National Laboratory, Argonne, Illinois
60439, USA}

\date{\today{}}
\begin{abstract}
The ability to tune exchange (magnetic) interactions between 3d transition metals in perovskite structures has proven to be a powerful route to discovery of novel properties. Here we demonstrate that the introduction of 3d-5d exchange pathways in double perovskites enables additional tunability, a result of the large spatial extent of 5d wave functions. Using x-ray probes of magnetism and structure at high pressure, we show that compression of $\mathrm{Sr_2FeOsO_6}$ drives an unexpected continuous change in the sign of Fe-Os exchange interactions and a transition from antiferromagnetic to ferrimagnetic order. We analyze the relevant electron-electron interactions, shedding light into fundamental differences with the more thoroughly studied 3d-3d systems.

 \end{abstract}

\maketitle

First-row (3d) transition metal oxides with perovskite crystal structure ($\mathrm{ABO_3}$ with A an alkali, alkaline earth or rare earth ion and B a transition metal ion) continue to provide a rich playground for the realization of novel quantum states, a result of a strong interplay between spin, orbital, charge and lattice degrees of freedom. Manipulation of electron correlations at interfaces of heterostructures or under application of electric and magnetic fields in multiferroic structures has unraveled a plethora of new phenomena in these strongly correlated 3d systems \cite{Huijben06, Ohtomo02, Cheong07,Tokunaga09, Ramesh07}. In the search for materials with additional tunability, double perovskites with $\mathrm{A_2BB'O_6}$ formula unit (B and B' are distinct TM ions arranged periodically, doubling the unit cell) have emerged as a new fertile ground for exploration of novel quantum states \cite{Saha13, Pardo09}. This is because d-orbital occupancy and symmetry, which together with lattice distortions control B-B' electron hoping integrals and therefore transport and exchange (magnetic) interactions, can be independently tuned at B and B' sites. Additionally, the ability to combine the rather localized 3d electron wave functions at B sites with the more delocalized 5d electron wavefunctions of third-row transition metal ions at B' sites, provides a path to further tunability and potential for new functionalities. 

While the mechanisms regulating 3d-3d exchange interactions in these oxide structures are rather well understood in terms of Goodenough-Kanamori (G-K) rules \cite{Anderson50, Goodenough55, Goodenough58, Kanamori59}, the understanding of 3d-5d interactions is much less developed. For example, on-site Coulomb interactions are significantly reduced at 5d sites relative to 3d sites, affecting the strength of Hund's coupling and related strength of double-exchange interactions (i.e., delocalized superexchange involving $e_g$ electrons)\cite{Zener51, Kanamori01, Anderson55}. Furthermore, the disparate crystal electric fields (CEF) at 3d and 5d sites, together with sizable spin-orbit interactions in 5d ions with strong nuclear potential alters the energy landscape and modifies electron hoping. As a result, the validity of G-K rules in 3d-5d systems ought to be addressed. The extended 5d orbitals may also require accounting for longer-range exchange pathways beyond first neighbor exchange in order to understand magnetic phenomena. A number of double perovskites of this type have shown remarkable  properties including half-metallicity above room temperature in $\mathrm{A_2CrWO_6}$ \cite{JBPhilipp03}, colossal magnetoresistance in $\mathrm{Sr_2FeReO_6}$ \cite{Kobayashi99}, and moderate to large magneto-optical properties in $\mathrm{Sr_2CrReO_6}$ and $\mathrm{Sr_2CrOsO_6}$ \cite{Hena08}, a testament to the exciting opportunities inherent in the exploration of these versatile structures.


\begin{figure}[t]

\includegraphics[scale=0.38]{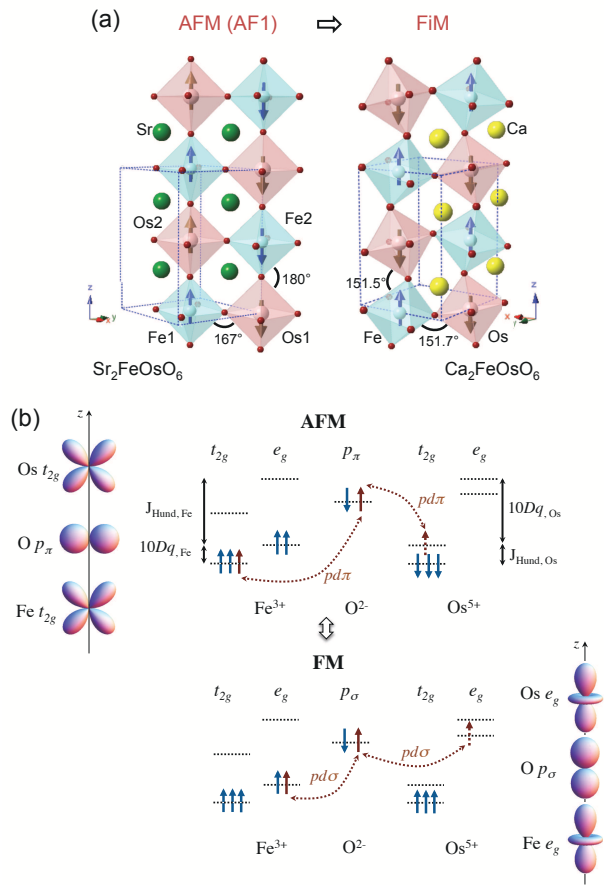}

\caption{(Color online) (a) Crystal and magnetic structure of $\mathrm{Sr_2FeOsO_6}$ and $\mathrm{Ca_2FeOsO_6}$ double perovskites. (b) Schematic of double exchange (FM) and superexchange (AFM) interactions between Fe$^{3+}$ and Os$^{5+}$ ions along c-axis.}
\label{fig:Fig1}
 
\end{figure}

In this work, we explore the use of external pressure (compressive stress) as a tool to modify lattice structure and 3d-5d exchange interactions, allowing us to develop a fundamental understanding of the underlying physics in a model system, $\mathrm{Sr_2FeOsO_6}$. Using x-ray spectroscopic and structural probes in a diamond anvil cell, we find a pressure-induced reversal in the sign of Fe-Os exchange interactions along the crystallographic c-axis of this tetragonal structure, driving a change from an antiferromagnetic (AFM) to a ferrimagnetic (FiM) ground state. The transition is a result of an increase in the difference between cubic crystal fields at Fe and Os sites with pressure, which gradually impairs the ability of $e_g$ electrons to mediate ferromagnetic (FM) double exchange (DE) interactions along the c-axis. As a result, AFM super exchange (SE) involving $t_{2g}$ electrons becomes the dominant interaction in both in-plane and out-of-plane directions at high pressure. This mechanism is different than that behind the AFM-FiM transition induced by chemical pressure (Ca substitution for Sr) where the weakening of DE-FM interactions is driven by an increased Fe-O-Os bond buckling along c-axis. That both mechanisms lead to a common ground state is striking evidence for a weakened FM interaction as a result of the delocalized nature of 5d wavefunctions, namely, a small on-site Coulomb interaction and large crystal field at Os sites. The novel pressure-induced transition, unique to the 3d-5d makeup of this double perovskite structure, transforms a material with no remanent magnetization or coercivity into one with robust coercivity ($\sim 0.5$ T) typical of permanent magnets, aided by the presence of spin-orbit interaction at Os sites. The fragility of FM interactions in this 3d-5d system indicates that modest changes in tensile or compressive strain in engineered epitaxial films could have significant impact on magnetic response providing a path to functional devices.

We start by comparing the structural and electronic properties of $\mathrm{Sr_{2}FeOsO_{6}}$ \cite{Feng13, Paul13} to its chemically compressed analog, $\mathrm{Ca_{2}FeOsO_{6}}$ \cite{Feng14}. The smaller Ca$^{2+}$ ions drive a transition from a tetragonal ($I4/m$) to a distorted monoclinic ($P2_1/n$) crystal structure (Fig. \hyperref[fig:Fig1] {1a}). In the tetragonal structure, the B/B'O$_6$ octahedra are rotated by $\mathrm{\sim}$ 13 degrees around the c-axis, leading to buckled ($\mathrm{\sim}$ 167 degrees) Fe-O-Os bonds within the ab plane but retaining Fe-O-Os collinearity along the c-axis. This collinear bonding displays FM coupling of Fe and Os ions, as expected from G-K rules for the coupling between Fe$^{3+}$ (3d$^5$) and Os$^{5+}$ (5d$^3$) ions \cite{Goodenough_book}. Surprisingly, the in-plane Fe-Os coupling is AFM despite the relatively small buckling angle, in apparent contradiction with G-K rules which predict AFM coupling at much larger angles \cite{Goodenough_book}. As discussed below, this is a result of weak 3d-5d FM interactions relative to their 3d-3d counterparts, driven by the delocalized nature of the 5d wavefunction. The compressed Ca structure adds a second B/B'O$_6$ octahedra rotation around [110] (a$^-$a$^-$b$^+$ in Glazer notation), leading to deviations from collinearity in both in-plane and out-of-plane directions and emergence of FiM order (Fig. \hyperref[fig:Fig1] {1a}). Note that the FM interaction between Fe and Os is mediated by the overlap of e$_g$ orbitals with oxygen p$_\sigma$ orbitals, highly sensitive to the degree of Fe-O-Os collinearity (Fig. \hyperref[fig:Fig1] {1b}, bottom panel). On the other hand, the AFM interaction is mediated by the overlap of t$_{2g}$ and oxygen p$_\pi$ orbitals, which is only weakly modified by buckling (see Supplemental Material \cite{SM} for a theoretical description).

\begin{figure}[t]
\centering
\includegraphics[scale=0.6]{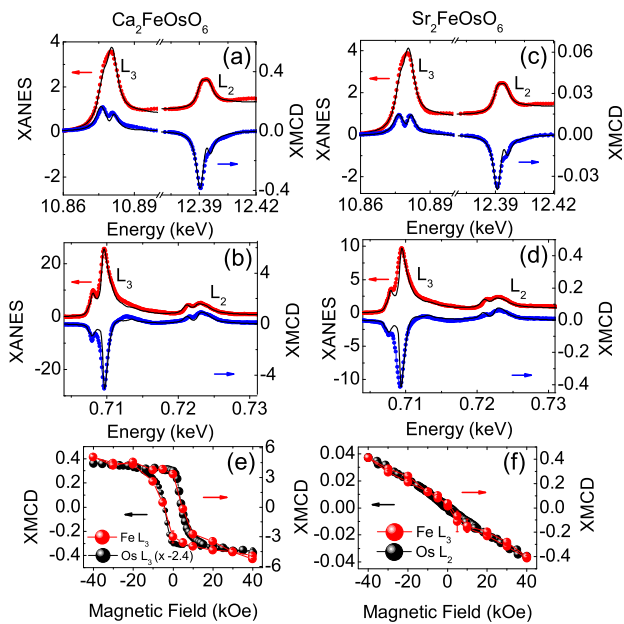}

\caption{(Color online) Normalized Os and Fe XANES and XMCD L$_{2,3}$ edge spectra for (a)-(b) $\mathrm{Ca_2FeOsO_6}$ and (c)-(d) $\mathrm{Sr_2FeOsO_6}$ double perovskites measured at ambient pressure. Black lines are results from cluster calculations. (e)-(f) Field-dependence of Fe L$_3$ (E$_{L_3}$=0.709 keV) and Os L$_{2,3}$ (E$_{L_2}$=12.391 keV and E$_{L_3}$=10.877 keV) XMCD peak intensities for $\mathrm{Ca_2FeOsO_6}$ and $\mathrm{Sr_2FeOsO_6}$.}
\label{fig:Fig2}
 
\end{figure}

X-ray magnetic circular dichroism (XMCD) measurements at Fe L$_{2,3}$ and Os L$_{2,3}$ edges confirm the presence of FiM order in the Ca structure and absence thereof in the Sr structure (Fig.~\ref{fig:Fig2}). The Fe and Os magnetic moments point in opposite directions as evident by the opposite signs of XMCD signals (Fe along applied field). A strong exchange coupling between magnetic sublattices is evident from their correlated magnetization reversal. The Sr structure shows typical AFM response to applied fields, namely, linear dependence of magnetization due to field-induced canting of Fe and Os moments. The low saturation magnetization in the Sr structure at 4 Tesla, ten times smaller than in the Ca structure (Fig. \hyperref[fig:Fig2]{2 e-f}), is indicative of strong exchange interactions favoring a nearly collinear AFM arrangement of magnetic moments with small canting.

Since spin-orbit interactions can affect electronic structure at 5d sites, we performed a theoretical analysis of the x-ray absorption data using OsO$_6$ and FeO$_6$ cluster calculations including single and double ligand-hole states \cite{SM}. A good agreement between the experimental and theoretical spectra is obtained (see Fig. \hyperref[fig:Fig2]{2 a-d}). Calculated orbital and spin moments for Os are $m_l$= $\mathrm{-\langle L_z \rangle }$ $= 0.27$ $ \mu_B$/Os, $m_s$= $-\mathrm{2\langle S_z \rangle}$$=-2.72$ $ \mu_B$/Os ($m_{\rm Os} = m_l+m_s = -2.45$ $\mu_B$/Os). The orbital magnetization at Os sites is opposite to the spin magnetization, as expected for a less than half-filled 5d orbital occupancy (calculated number of 5d holes is $n_h$=6.38). The corresponding quantities for Fe sites are $m_l = 0.044$ $\mu_B$/Fe, $m_s = 4.34$ $\mu_B$/Fe ($m_{Fe} = 4.38$ $\mu_B$/Fe; number of 3d holes $n_h$=4.35). The XMCD-derived orbital and spin moments as well as net saturated magnetization (m$_{Fe}$ + m$_{Os}$ = 1.93 $\mu_B$) are in good agreement with DFT calculations \cite{Hongbo14} and SQUID measurements \cite{Feng13}. The results confirm a high-spin ground state for Fe, with covalency (and S-O interactions at Os sites) responsible for the reduction in moment values from the expected 5(3)$\mu_B$ of Fe$^{3+}$ (Os$^{5+}$) ions in octahedral CEF (neutron diffraction experiments have so far failed to converge on a consistent description of the magnitude of local moments in either structure \cite{Paul_PRL13,Morrow14}). The orbital-to-spin moment ratio is ten times larger at Os sites due to the stronger S-O interaction. Analysis of the Os L$_{2,3}$ branching ratio \cite{SM} indicates that Os cannot be described in the strong S-O coupling limit commonly used for Iridates \cite{Kim08,Haskel12,Laguna-Marco10} and instead falls in an intermediate regime where S-O interactions compete with Hund's exchange in the presence of a dominant CEF interaction in agreement with calculations \cite{Hongbo14}. S-O and CEF interactions acting on Os 5d electrons are expected to be a source of magnetic anisotropy and explain the sizable coercivity of FiM Ca$_2$FeOsO$_6$.

\begin{figure}[b]
\centering
\includegraphics[scale=0.48]{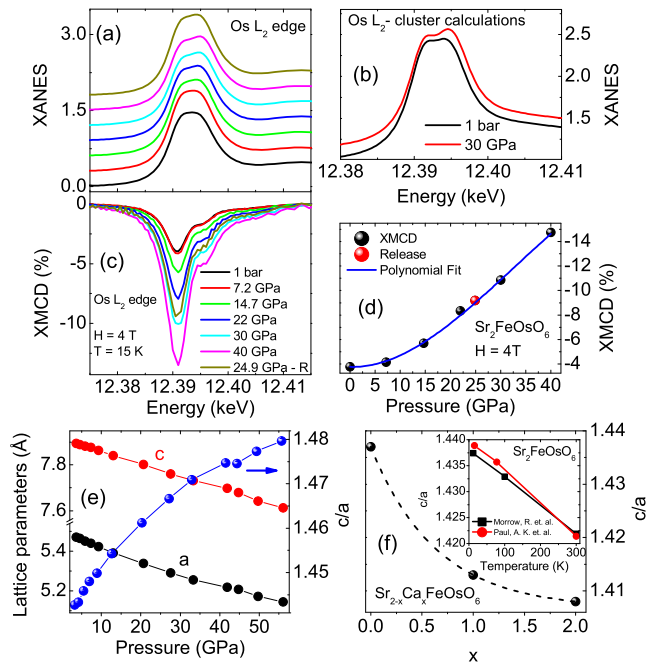}

\caption{\label{fig:Fig3} (Color online) (a) Pressure dependence of Os L$_2$-edge XANES for $\mathrm{Sr_2FeOsO_6}$. (b) XANES spectra obtained from cluster calculations using octahedral crystal field values of 10$Dq$ = 3.7 eV and 4.2 eV for ambient pressure and 30 GPa, respectively. (c)-(d) Os L$_2$-edge XMCD as a function of applied pressure. Data collected after partial pressure release (24.9 GPa) are also shown. (e) Tetragonal lattice parameters and c/a ratio refined within I4/m tetragonal space group. (f) c/a ratio as a function of Ca doping in $\mathrm{Sr_{2-x}Ca_xFeOsO_6}$, and as a function of temperature (ambient pressure) in $\mathrm{Sr_2FeOsO_6}$, reproduced from Ref.~\cite{Paul13, Morrow14}. } 
\end{figure}

\begin{figure}[t]
\centering
\includegraphics[scale=0.35]{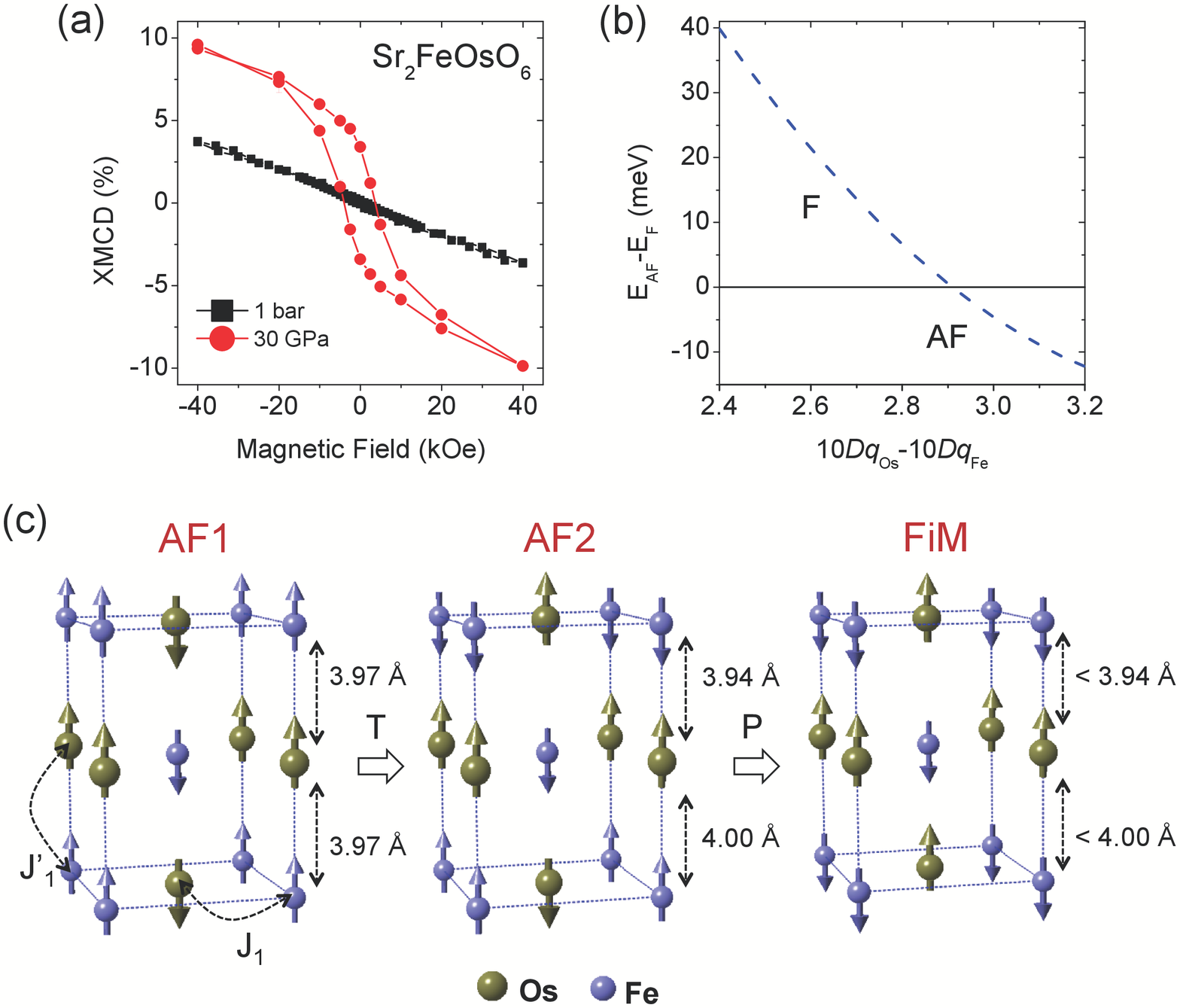}

\caption{(a) Field-dependence of Os L$_2$-edge XMCD intensity at ambient pressure and at 30 GPa showing emergence of FM response in the Os sub lattice. (b) Energy difference between FM and AFM Fe-Os coupling along the c-axis as a function of the difference in crystal electric field at Os and Fe sites. (c) Evolution of magnetic structure with cooling and applied pressure. J$_1$ (J$_1^{\prime}$) denotes nearest-neighbor exchange between Fe and Os moments. The Fe-Os distances in AF1 and AF2 structures are taken from Ref \cite{Paul_PRL13}.}
\label{fig:Fig4}
 
\end{figure}

We now turn to the effect of hydrostatic pressure on the structural and electronic ground state of $\mathrm{Sr_2FeOsO_6}$. Experimental details for the XRD and XMCD measurements carried out in diamond anvil cells at low temperature are given in the supplemental material \cite{SM}. As seen in Fig.~\hyperref[fig:Fig3]{3 c-d}, a continuous increase in XMCD signal is observed under pressure reaching four times its ambient pressure value at 40 GPa (although not matching the saturation value of the Ca structure). The emergence of a FM response in the Os sublattice is clearly seen in the field-dependent XMCD data, where large coercivity (0.5 Tesla) comparable to that of the Ca structure (0.85 Tesla) is observed (Fig. \hyperref[fig:Fig4]{4a}). Remanent magnetization ($\sim$ 0.23 $\mu_B$/Os) also emerges. This presents a dramatic change from the negligible coercivity and remanence measured at ambient pressure (Fig. \hyperref[fig:Fig4]{4a}). We note that the XMCD signal fully reverts in size upon pressure reduction from 40 GPa to 24.9 GPa with no measurable hysteresis. It appears that the exchange coupling of Os and Fe moments along the c-axis continuously transforms from FM towards AFM under lattice compression leading to a FiM response, mimicking the behavior of the chemically compressed Ca structure (Fig. \hyperref[fig:Fig1]{1a}). 

One may be tempted to conclude that the driving force for the magnetic transition is a pressure-induced monoclinic distortion with related deviation in Fe-O-Os c-axis bonding from collinearity. However, our low temperature (15 K) x-ray powder diffraction measurements show that the Sr structure remains tetragonal to 56 GPa \cite{SM}. Ca doping drives a continuous {\em reduction} in c/a ratio, a result of larger buckling in c-axis Fe-O-Os bonding compressing the c-axis lattice parameter faster than the a-axis lattice parameter (Fig.~\hyperref[fig:Fig3]{3f}). This contrasts with the {\em increase} in c/a ratio with pressure (Fig. \hyperref[fig:Fig3]{3e}), which indicates that such c-axis buckling does not take place in the Sr structure. Indeed, an enhanced c/a ratio is also observed upon cooling the Sr structure at ambient pressure (inset of Fig.~\hyperref[fig:Fig3]{3f}), which is known to retain tetragonal symmetry with increased in-plane Fe-O-Os buckling. X-ray absorption fine structure (XAFS) data at the Os L$_2$ edge shows no significant change to 40 GPa \cite{SM}. Since XAFS is highly sensitive to deviations in Os-O-Fe bonding from collinearity \cite{SM} we conclude that this bonding remains collinear along the c-axis and therefore the transition to a FiM state under pressure has a different origin than that induced by chemical pressure.

Experimental and theoretical reports on the nature of exchange interactions in $\mathrm{Sr_2FeOsO_6}$ have addressed the origin of AFM order in this structure at ambient pressure \cite{Paul_PRL13,Morrow14, Kanungo14}. In fact, two different AFM configurations (AF1 and AF2, Fig. \hyperref[fig:Fig4]{4c}) are observed by neutron diffraction to coexist at low temperature, the AF2 phase appearing on cooling below 67 K and becoming dominant below 55 K (85$\%$ at 2 K) \cite{Paul_PRL13,Morrow14}. The AF1 phase consists of FM Fe-Os chains along the c-axis, while half of Fe-Os bonds become AFM coupled in the AF2 phase as a result of a lattice distortion in which short (AFM) and long (FM) Fe-Os distances are created. 

There are two direct consequences of the reduction in the Fe-O-Os bond lengths. First, there is an increase in the hoping integrals between the TM ions and the oxygens. These are given by the tight-binding parameters $pd\sigma_{TM-O}$ and $pd\pi_{TM-O}$ for the $\sigma$-bonding $e_g$ orbitals and $\pi$-bonding t$_{2g}$ orbitals, respectively. Both hoping matrix elements have a $r_{\rm TM-O}^{-3.5}$ dependence on the TM-O distance. Additionally, both the SE and DE interactions have, to leading order, a similar dependence on the hoping integrals \cite{SM}. Therefore, changes in the relative strengths of the DE and SE interactions are a higher-order effect. This is confirmed by numerical calculations using Fe-O-Os clusters with parameters consistent with those used to calculate the x-ray absorption spectra, which showed a minimal dependence of the {\em relative} strengths of the exchange interactions with the hoping integrals. The second quantity affected by the reduction of the TM-O distance is the CEF. Here, the magnetic exchange interactions are affected differently. The DE involves the exchange of an e$_g$ electron between the Fe and the Os ion. To leading order the gain in energy due to the DE is given by

\begin{equation}
\Delta E_{DE}  \approx -\frac{(pd\sigma)^4}{\Delta^2} \frac {J_{Hund, Os}}{(U+10Dq_{Os}-10Dq_{Fe})^2}
\end{equation}

where $\Delta$ is the charge-transfer energy for osmium and oxygen, $U$ is the charge transfer from the Fe site to the Os site \cite{SM}. We see that the DE is directly proportional to the (weak) Hund's rule coupling $J_{Hund,Os}$ on the osmium ion. A decrease in lattice parameters causes an increase in the difference between the Os and Fe crystal-field parameters $10Dq_{Os}-10Dq_{Fe}$, leading to a further reduction of the double exchange interaction. This has been confirmed by cluster calculations where a transition from a FM to AFM Fe-Os coupling is obtained (Fig.~\hyperref[fig:Fig4]{4b}). Evidence for pressure-induced changes in CEF parameter at Os sites is seen in the evolution of the L$_2$-edge white line (Fig. \hyperref[fig:Fig3]{3a}), the enhanced spectral weight in the high-energy side reproduced in cluster calculations with a 0.5 eV increase in crystal field strength (Fig.~\hyperref[fig:Fig3]{3b}). 

An understanding of the pressure induced magnetic transition naturally follows. Upon lattice compression the reduction in all Fe-Os distances and related increase in (cubic) CEF energy causes a weakening of an already fragile Fe-Os DE-FM interaction relative to the SE-AFM interaction along the c-axis. The in-plane Fe-Os coupling remains AFM at all temperatures and pressures due to the persistent in-plane Fe-O-Os buckling. The transition to AFM coupling in Fe-Os bonds along the c-axis leads to emergence of FiM order in the compressed structure. Although full, collinear FiM order is not achieved at 40 GPa (based on field dependent data and comparison to Ca analog), the transition appears to be second order as no detectable hysteresis was observed on pressure release. 

The high degree of tunability of exchange interactions under pressure is unique to the 3d-5d makeup of this double-perovskite structure, namely, the delocalized 5d state reduces the onsite Coulomb interaction and increases CEF resulting in weaker DE-FM Fe-Os interactions compared to those found in 3d-3d analogs. For example, the FM state of $\mathrm{La_2MnNiO_6}$ remains unchanged to at least 38 GPa, a result of strong on-site (Hund's) coupling on Mn sites \cite{Haskel11}. The ability to not only dramatically alter the magnetic state with external stimuli but to do so in a continuous fashion is appealing for applications where tunability of coercivity and saturation magnetization is desired. While undoped $\mathrm{Sr_2FeOsO_6}$ requires sizable pressures ($>$ 10 GPa) to induce significant changes in coercivity and net magnetization, one can envision doping Ca into the Sr structure to attain proximity to a FiM state \cite{Morrow14} hence enabling much smaller pressures (or strain in films) to drive the AFM-FiM transition in a reversible way; e.g., with dynamical compression/decompression. Furthermore, the ability to achieve sizable coercive fields (0.5 Tesla) by manipulating indirect (3d-5d) exchange interactions with pressure coupled with S-O interactions at 5d sites presents an interesting opportunity in the search for rare-earth free permanent magnets \cite{Zeng02,Balamurugan14}.

\begin{acknowledgments}
Work at Argonne is supported by the US Department of Energy, Office of Science, Office of Basic Energy Sciences, under Contract No. DE-AC- 02-06CH11357. L. S. I. Veiga is supported by FAPESP (SP-Brazil) under Contract No. 2013/14338-3. M. v. V was supported by DOE-BES under Award No. DEFG02-03ER46097. This research was supported in part by the World Premier International Research Center of the Ministry of Education, Culture, Sports, Science and Technology (MEXT) of Japan, the Japan Society for the Promotion of Science (JSPS) through a Grant-in-Aid for Scientific Research (25289233). We would like to thank Changyong Park and Curtis Kenny-Benson for their assistance at 16-BM-D beamline, Sergey N. Tkachev for assistance with gas loading, and Richard Rosenberg for help with soft x-ray measurements. We also thank GSECARS for use of the gas loading and laser drilling facilities.
\end{acknowledgments}

\bibliographystyle{apsrev}

\bibliography{references}

\begin{thebibliography}{32}
\expandafter\ifx\csname natexlab\endcsname\relax\def\natexlab#1{#1}\fi
\expandafter\ifx\csname bibnamefont\endcsname\relax
  \def\bibnamefont#1{#1}\fi
\expandafter\ifx\csname bibfnamefont\endcsname\relax
  \def\bibfnamefont#1{#1}\fi
\expandafter\ifx\csname citenamefont\endcsname\relax
  \def\citenamefont#1{#1}\fi
\expandafter\ifx\csname url\endcsname\relax
  \def\url#1{\texttt{#1}}\fi
\expandafter\ifx\csname urlprefix\endcsname\relax\def\urlprefix{URL }\fi
\providecommand{\bibinfo}[2]{#2}
\providecommand{\eprint}[2][]{\url{#2}}

\bibitem[{\citenamefont{Huijben et~al.}(2006)\citenamefont{Huijben, Rijnders,
  Blank, Bals, Aert, Verbeeck, Tendeloo, Brinkman, and Hilgenkamp}}]{Huijben06}
\bibinfo{author}{\bibfnamefont{M.}~\bibnamefont{Huijben}},
  \bibinfo{author}{\bibfnamefont{G.}~\bibnamefont{Rijnders}},
  \bibinfo{author}{\bibfnamefont{D.~H.~A.} \bibnamefont{Blank}},
  \bibinfo{author}{\bibfnamefont{S.}~\bibnamefont{Bals}},
  \bibinfo{author}{\bibfnamefont{S.~V.} \bibnamefont{Aert}},
  \bibinfo{author}{\bibfnamefont{J.}~\bibnamefont{Verbeeck}},
  \bibinfo{author}{\bibfnamefont{G.~V.} \bibnamefont{Tendeloo}},
  \bibinfo{author}{\bibfnamefont{A.}~\bibnamefont{Brinkman}}, \bibnamefont{and}
  \bibinfo{author}{\bibfnamefont{H.}~\bibnamefont{Hilgenkamp}},
  \bibinfo{journal}{Nat Mater} \textbf{\bibinfo{volume}{5}},
  \bibinfo{pages}{556} (\bibinfo{year}{2006}).

\bibitem[{\citenamefont{Ohtomo et~al.}(2002)\citenamefont{Ohtomo, Muller,
  Grazul, and Hwang}}]{Ohtomo02}
\bibinfo{author}{\bibfnamefont{A.}~\bibnamefont{Ohtomo}},
  \bibinfo{author}{\bibfnamefont{D.~A.} \bibnamefont{Muller}},
  \bibinfo{author}{\bibfnamefont{J.~L.} \bibnamefont{Grazul}},
  \bibnamefont{and} \bibinfo{author}{\bibfnamefont{H.~Y.} \bibnamefont{Hwang}},
  \bibinfo{journal}{Nature} \textbf{\bibinfo{volume}{419}},
  \bibinfo{pages}{378} (\bibinfo{year}{2002}).

\bibitem[{\citenamefont{Cheong and Mostovoy}(2007)}]{Cheong07}
\bibinfo{author}{\bibfnamefont{S.-W.} \bibnamefont{Cheong}} \bibnamefont{and}
  \bibinfo{author}{\bibfnamefont{M.}~\bibnamefont{Mostovoy}},
  \bibinfo{journal}{Nat Mater} \textbf{\bibinfo{volume}{6}},
  \bibinfo{pages}{13} (\bibinfo{year}{2007}).

\bibitem[{\citenamefont{Tokunaga et~al.}(2009)\citenamefont{Tokunaga, Furukawa,
  Sakai, Taguchi, Arima, and Tokura}}]{Tokunaga09}
\bibinfo{author}{\bibfnamefont{Y.}~\bibnamefont{Tokunaga}},
  \bibinfo{author}{\bibfnamefont{N.}~\bibnamefont{Furukawa}},
  \bibinfo{author}{\bibfnamefont{H.}~\bibnamefont{Sakai}},
  \bibinfo{author}{\bibfnamefont{Y.}~\bibnamefont{Taguchi}},
  \bibinfo{author}{\bibfnamefont{T.-h.} \bibnamefont{Arima}}, \bibnamefont{and}
  \bibinfo{author}{\bibfnamefont{Y.}~\bibnamefont{Tokura}},
  \bibinfo{journal}{Nat Mater} \textbf{\bibinfo{volume}{8}},
  \bibinfo{pages}{558} (\bibinfo{year}{2009}).

\bibitem[{\citenamefont{Ramesh and Spaldin}(2007)}]{Ramesh07}
\bibinfo{author}{\bibfnamefont{R.}~\bibnamefont{Ramesh}} \bibnamefont{and}
  \bibinfo{author}{\bibfnamefont{N.~A.} \bibnamefont{Spaldin}},
  \bibinfo{journal}{Nat Mater} \textbf{\bibinfo{volume}{6}},
  \bibinfo{pages}{21} (\bibinfo{year}{2007}).

\bibitem[{\citenamefont{Saha-Dasgupta}(2013)}]{Saha13}
\bibinfo{author}{\bibfnamefont{T.}~\bibnamefont{Saha-Dasgupta}},
  \bibinfo{journal}{J. Supercond. Nov. Magn.} \textbf{\bibinfo{volume}{26}},
  \bibinfo{pages}{1991} (\bibinfo{year}{2013}).

\bibitem[{\citenamefont{Pardo and Pickett}(2009)}]{Pardo09}
\bibinfo{author}{\bibfnamefont{V.}~\bibnamefont{Pardo}} \bibnamefont{and}
  \bibinfo{author}{\bibfnamefont{W.~E.} \bibnamefont{Pickett}},
  \bibinfo{journal}{Phys. Rev. B} \textbf{\bibinfo{volume}{80}},
  \bibinfo{pages}{054415} (\bibinfo{year}{2009}).

\bibitem[{\citenamefont{Anderson}(1950)}]{Anderson50}
\bibinfo{author}{\bibfnamefont{P.~W.} \bibnamefont{Anderson}},
  \bibinfo{journal}{Phys. Rev.} \textbf{\bibinfo{volume}{79}},
  \bibinfo{pages}{350} (\bibinfo{year}{1950}).

\bibitem[{\citenamefont{Goodenough}(1955)}]{Goodenough55}
\bibinfo{author}{\bibfnamefont{J.~B.} \bibnamefont{Goodenough}},
  \bibinfo{journal}{Phys. Rev.} \textbf{\bibinfo{volume}{100}},
  \bibinfo{pages}{564} (\bibinfo{year}{1955}).

\bibitem[{\citenamefont{Goodenough}(1958)}]{Goodenough58}
\bibinfo{author}{\bibfnamefont{J.~B.} \bibnamefont{Goodenough}},
  \bibinfo{journal}{J. Phys. Chem. Solids} \textbf{\bibinfo{volume}{6}},
  \bibinfo{pages}{287} (\bibinfo{year}{1958}).

\bibitem[{\citenamefont{Kanamori}(1959)}]{Kanamori59}
\bibinfo{author}{\bibfnamefont{J.}~\bibnamefont{Kanamori}},
  \bibinfo{journal}{J. Phys. Chem. Solids} \textbf{\bibinfo{volume}{10}},
  \bibinfo{pages}{87} (\bibinfo{year}{1959}).

\bibitem[{\citenamefont{Zener}(1951)}]{Zener51}
\bibinfo{author}{\bibfnamefont{C.}~\bibnamefont{Zener}},
  \bibinfo{journal}{Phys. Rev.} \textbf{\bibinfo{volume}{82}},
  \bibinfo{pages}{403} (\bibinfo{year}{1951}).

\bibitem[{\citenamefont{Kanamori and Terakura}(2001)}]{Kanamori01}
\bibinfo{author}{\bibfnamefont{J.}~\bibnamefont{Kanamori}} \bibnamefont{and}
  \bibinfo{author}{\bibfnamefont{K.}~\bibnamefont{Terakura}},
  \bibinfo{journal}{J. Phys. Soc. Jpn.} \textbf{\bibinfo{volume}{70}},
  \bibinfo{pages}{1433} (\bibinfo{year}{2001}).

\bibitem[{\citenamefont{Anderson and Hasegawa}(1955)}]{Anderson55}
\bibinfo{author}{\bibfnamefont{P.}~\bibnamefont{Anderson}} \bibnamefont{and}
  \bibinfo{author}{\bibfnamefont{H.}~\bibnamefont{Hasegawa}},
  \bibinfo{journal}{Phys. Rev.} \textbf{\bibinfo{volume}{100}},
  \bibinfo{pages}{675} (\bibinfo{year}{1955}).

\bibitem[{\citenamefont{Philipp et~al.}(2003)\citenamefont{Philipp, Majewski,
  Alff, Erb, Gross, Graf, Brandt, Simon, Walther, Mader et~al.}}]{JBPhilipp03}
\bibinfo{author}{\bibfnamefont{J.~B.} \bibnamefont{Philipp}},
  \bibinfo{author}{\bibfnamefont{P.}~\bibnamefont{Majewski}},
  \bibinfo{author}{\bibfnamefont{L.}~\bibnamefont{Alff}},
  \bibinfo{author}{\bibfnamefont{A.}~\bibnamefont{Erb}},
  \bibinfo{author}{\bibfnamefont{R.}~\bibnamefont{Gross}},
  \bibinfo{author}{\bibfnamefont{T.}~\bibnamefont{Graf}},
  \bibinfo{author}{\bibfnamefont{M.~S.} \bibnamefont{Brandt}},
  \bibinfo{author}{\bibfnamefont{J.}~\bibnamefont{Simon}},
  \bibinfo{author}{\bibfnamefont{T.}~\bibnamefont{Walther}},
  \bibinfo{author}{\bibfnamefont{W.}~\bibnamefont{Mader}},
  \bibnamefont{et~al.}, \bibinfo{journal}{Phys. Rev. B}
  \textbf{\bibinfo{volume}{68}}, \bibinfo{pages}{144431}
  (\bibinfo{year}{2003}).

\bibitem[{\citenamefont{Kobayashi et~al.}(1999)\citenamefont{Kobayashi, Kimura,
  Tomioka, Sawada, Terakura, and Tokura}}]{Kobayashi99}
\bibinfo{author}{\bibfnamefont{K.-I.} \bibnamefont{Kobayashi}},
  \bibinfo{author}{\bibfnamefont{T.}~\bibnamefont{Kimura}},
  \bibinfo{author}{\bibfnamefont{Y.}~\bibnamefont{Tomioka}},
  \bibinfo{author}{\bibfnamefont{H.}~\bibnamefont{Sawada}},
  \bibinfo{author}{\bibfnamefont{K.}~\bibnamefont{Terakura}}, \bibnamefont{and}
  \bibinfo{author}{\bibfnamefont{Y.}~\bibnamefont{Tokura}},
  \bibinfo{journal}{Phys. Rev. B} \textbf{\bibinfo{volume}{59}},
  \bibinfo{pages}{11159} (\bibinfo{year}{1999}).

\bibitem[{\citenamefont{Das et~al.}(2008)\citenamefont{Das, De~Raychaudhury,
  and Saha-Dasgupta}}]{Hena08}
\bibinfo{author}{\bibfnamefont{H.}~\bibnamefont{Das}},
  \bibinfo{author}{\bibfnamefont{M.}~\bibnamefont{De~Raychaudhury}},
  \bibnamefont{and}
  \bibinfo{author}{\bibfnamefont{T.}~\bibnamefont{Saha-Dasgupta}},
  \bibinfo{journal}{Applied Physics Letters} \textbf{\bibinfo{volume}{92}},
  (\bibinfo{year}{2008}).

\bibitem[{\citenamefont{Feng et~al.}(2013)\citenamefont{Feng, Tsujimoto, Guo,
  Sun, Sathish, and Yamaura}}]{Feng13}
\bibinfo{author}{\bibfnamefont{H.~L.} \bibnamefont{Feng}},
  \bibinfo{author}{\bibfnamefont{Y.}~\bibnamefont{Tsujimoto}},
  \bibinfo{author}{\bibfnamefont{Y.}~\bibnamefont{Guo}},
  \bibinfo{author}{\bibfnamefont{Y.}~\bibnamefont{Sun}},
  \bibinfo{author}{\bibfnamefont{C.~I.} \bibnamefont{Sathish}},
  \bibnamefont{and} \bibinfo{author}{\bibfnamefont{K.}~\bibnamefont{Yamaura}},
  \bibinfo{journal}{High Pressure Res.} \textbf{\bibinfo{volume}{33}},
  \bibinfo{pages}{221} (\bibinfo{year}{2013}).

\bibitem[{\citenamefont{Paul et~al.}(2013{\natexlab{a}})\citenamefont{Paul,
  Jansen, Yan, Felser, Reehuis, and Abdala}}]{Paul13}
\bibinfo{author}{\bibfnamefont{A.~K.} \bibnamefont{Paul}},
  \bibinfo{author}{\bibfnamefont{M.}~\bibnamefont{Jansen}},
  \bibinfo{author}{\bibfnamefont{B.}~\bibnamefont{Yan}},
  \bibinfo{author}{\bibfnamefont{C.}~\bibnamefont{Felser}},
  \bibinfo{author}{\bibfnamefont{M.}~\bibnamefont{Reehuis}}, \bibnamefont{and}
  \bibinfo{author}{\bibfnamefont{P.~M.} \bibnamefont{Abdala}},
  \bibinfo{journal}{Inorganic Chemistry} \textbf{\bibinfo{volume}{52}},
  \bibinfo{pages}{6713} (\bibinfo{year}{2013}{\natexlab{a}}).

\bibitem[{\citenamefont{Feng et~al.}(2014)\citenamefont{Feng, Arai, Matsushita,
  Tsujimoto, Guo, Sathish, Wang, Yuan, Tanaka, and Yamaura}}]{Feng14}
\bibinfo{author}{\bibfnamefont{H.~L.} \bibnamefont{Feng}},
  \bibinfo{author}{\bibfnamefont{M.}~\bibnamefont{Arai}},
  \bibinfo{author}{\bibfnamefont{Y.}~\bibnamefont{Matsushita}},
  \bibinfo{author}{\bibfnamefont{Y.}~\bibnamefont{Tsujimoto}},
  \bibinfo{author}{\bibfnamefont{Y.}~\bibnamefont{Guo}},
  \bibinfo{author}{\bibfnamefont{C.~I.} \bibnamefont{Sathish}},
  \bibinfo{author}{\bibfnamefont{X.}~\bibnamefont{Wang}},
  \bibinfo{author}{\bibfnamefont{Y.-H.} \bibnamefont{Yuan}},
  \bibinfo{author}{\bibfnamefont{M.}~\bibnamefont{Tanaka}}, \bibnamefont{and}
  \bibinfo{author}{\bibfnamefont{K.}~\bibnamefont{Yamaura}},
  \bibinfo{journal}{J. Am. Chem. Soc.} \textbf{\bibinfo{volume}{136}},
  \bibinfo{pages}{3326} (\bibinfo{year}{2014}).

\bibitem[{\citenamefont{Goodenough}(1963)}]{Goodenough_book}
\bibinfo{author}{\bibfnamefont{J.~B.} \bibnamefont{Goodenough}},
  \emph{\bibinfo{title}{Magnetism and the Chemical Bond}}
  (\bibinfo{publisher}{Interscience}, \bibinfo{address}{New York},
  \bibinfo{year}{1963}).

\bibitem[{SM()}]{SM}
\emph{\bibinfo{title}{See supplemental material for further details}}.

\bibitem[{\citenamefont{Wang et~al.}(2014)\citenamefont{Wang, Zhu, Ou, and
  Wu}}]{Hongbo14}
\bibinfo{author}{\bibfnamefont{H.}~\bibnamefont{Wang}},
  \bibinfo{author}{\bibfnamefont{S.}~\bibnamefont{Zhu}},
  \bibinfo{author}{\bibfnamefont{X.}~\bibnamefont{Ou}}, \bibnamefont{and}
  \bibinfo{author}{\bibfnamefont{H.}~\bibnamefont{Wu}}, \bibinfo{journal}{Phys.
  Rev. B} \textbf{\bibinfo{volume}{90}}, \bibinfo{pages}{054406}
  (\bibinfo{year}{2014}).

\bibitem[{\citenamefont{Paul et~al.}(2013{\natexlab{b}})\citenamefont{Paul,
  Reehuis, Ksenofontov, Yan, Hoser, T\"obbens, Abdala, Adler, Jansen, and
  Felser}}]{Paul_PRL13}
\bibinfo{author}{\bibfnamefont{A.~K.} \bibnamefont{Paul}},
  \bibinfo{author}{\bibfnamefont{M.}~\bibnamefont{Reehuis}},
  \bibinfo{author}{\bibfnamefont{V.}~\bibnamefont{Ksenofontov}},
  \bibinfo{author}{\bibfnamefont{B.}~\bibnamefont{Yan}},
  \bibinfo{author}{\bibfnamefont{A.}~\bibnamefont{Hoser}},
  \bibinfo{author}{\bibfnamefont{D.~M.} \bibnamefont{T\"obbens}},
  \bibinfo{author}{\bibfnamefont{P.~M.} \bibnamefont{Abdala}},
  \bibinfo{author}{\bibfnamefont{P.}~\bibnamefont{Adler}},
  \bibinfo{author}{\bibfnamefont{M.}~\bibnamefont{Jansen}}, \bibnamefont{and}
  \bibinfo{author}{\bibfnamefont{C.}~\bibnamefont{Felser}},
  \bibinfo{journal}{Phys. Rev. Lett.} \textbf{\bibinfo{volume}{111}},
  \bibinfo{pages}{167205} (\bibinfo{year}{2013}{\natexlab{b}}).

\bibitem[{\citenamefont{Morrow et~al.}(2014)\citenamefont{Morrow, Freeland, and
  Woodward}}]{Morrow14}
\bibinfo{author}{\bibfnamefont{R.}~\bibnamefont{Morrow}},
  \bibinfo{author}{\bibfnamefont{J.~F.} \bibnamefont{Freeland}},
  \bibnamefont{and} \bibinfo{author}{\bibfnamefont{P.}~\bibnamefont{Woodward}},
  \bibinfo{journal}{Inorg. Chem.} \textbf{\bibinfo{volume}{53}},
  \bibinfo{pages}{7983} (\bibinfo{year}{2014}).

\bibitem[{\citenamefont{Kim et~al.}(2008)\citenamefont{Kim, Ohsumi, Komesu,
  Sakai, Morita, Takagi, and Arima}}]{Kim08}
\bibinfo{author}{\bibfnamefont{B.~J.} \bibnamefont{Kim}},
  \bibinfo{author}{\bibfnamefont{H.}~\bibnamefont{Ohsumi}},
  \bibinfo{author}{\bibfnamefont{T.}~\bibnamefont{Komesu}},
  \bibinfo{author}{\bibfnamefont{S.}~\bibnamefont{Sakai}},
  \bibinfo{author}{\bibfnamefont{T.}~\bibnamefont{Morita}},
  \bibinfo{author}{\bibfnamefont{H.}~\bibnamefont{Takagi}}, \bibnamefont{and}
  \bibinfo{author}{\bibfnamefont{T.}~\bibnamefont{Arima}},
  \bibinfo{journal}{Science} \textbf{\bibinfo{volume}{323}},
  \bibinfo{pages}{1329} (\bibinfo{year}{2008}).

\bibitem[{\citenamefont{Haskel et~al.}(2012)\citenamefont{Haskel, Fabbris,
  Zhernenkov, Kong, Jin, Cao, and van Veenendaal}}]{Haskel12}
\bibinfo{author}{\bibfnamefont{D.}~\bibnamefont{Haskel}},
  \bibinfo{author}{\bibfnamefont{G.}~\bibnamefont{Fabbris}},
  \bibinfo{author}{\bibfnamefont{M.}~\bibnamefont{Zhernenkov}},
  \bibinfo{author}{\bibfnamefont{P.~P.} \bibnamefont{Kong}},
  \bibinfo{author}{\bibfnamefont{C.~Q.} \bibnamefont{Jin}},
  \bibinfo{author}{\bibfnamefont{G.}~\bibnamefont{Cao}}, \bibnamefont{and}
  \bibinfo{author}{\bibfnamefont{M.}~\bibnamefont{van Veenendaal}},
  \bibinfo{journal}{Phys. Rev. Lett.} \textbf{\bibinfo{volume}{109}},
  \bibinfo{pages}{027204} (\bibinfo{year}{2012}).

\bibitem[{\citenamefont{Laguna-Marco et~al.}(2010)\citenamefont{Laguna-Marco,
  Haskel, Souza-Neto, Lang, Krishnamurthy, Chikara, Cao, and van
  Veenendaal}}]{Laguna-Marco10}
\bibinfo{author}{\bibfnamefont{M.~A.} \bibnamefont{Laguna-Marco}},
  \bibinfo{author}{\bibfnamefont{D.}~\bibnamefont{Haskel}},
  \bibinfo{author}{\bibfnamefont{N.}~\bibnamefont{Souza-Neto}},
  \bibinfo{author}{\bibfnamefont{J.~C.} \bibnamefont{Lang}},
  \bibinfo{author}{\bibfnamefont{V.~V.} \bibnamefont{Krishnamurthy}},
  \bibinfo{author}{\bibfnamefont{S.}~\bibnamefont{Chikara}},
  \bibinfo{author}{\bibfnamefont{G.}~\bibnamefont{Cao}}, \bibnamefont{and}
  \bibinfo{author}{\bibfnamefont{M.}~\bibnamefont{van Veenendaal}},
  \bibinfo{journal}{Phys. Rev. Lett.} \textbf{\bibinfo{volume}{105}},
  \bibinfo{pages}{216407} (\bibinfo{year}{2010}).

\bibitem[{\citenamefont{Kanungo et~al.}(2014)\citenamefont{Kanungo, Yan,
  Jansen, and Felser}}]{Kanungo14}
\bibinfo{author}{\bibfnamefont{S.}~\bibnamefont{Kanungo}},
  \bibinfo{author}{\bibfnamefont{B.}~\bibnamefont{Yan}},
  \bibinfo{author}{\bibfnamefont{M.}~\bibnamefont{Jansen}}, \bibnamefont{and}
  \bibinfo{author}{\bibfnamefont{C.}~\bibnamefont{Felser}},
  \bibinfo{journal}{Phys. Rev. B} \textbf{\bibinfo{volume}{89}},
  \bibinfo{pages}{214414} (\bibinfo{year}{2014}).

\bibitem[{\citenamefont{Haskel et~al.}(2011)\citenamefont{Haskel, Fabbris,
  Souza-Neto, van Veenendaal, Shen, Smith, and Subramanian}}]{Haskel11}
\bibinfo{author}{\bibfnamefont{D.}~\bibnamefont{Haskel}},
  \bibinfo{author}{\bibfnamefont{G.}~\bibnamefont{Fabbris}},
  \bibinfo{author}{\bibfnamefont{N.~M.} \bibnamefont{Souza-Neto}},
  \bibinfo{author}{\bibfnamefont{M.}~\bibnamefont{van Veenendaal}},
  \bibinfo{author}{\bibfnamefont{G.}~\bibnamefont{Shen}},
  \bibinfo{author}{\bibfnamefont{A.~E.} \bibnamefont{Smith}}, \bibnamefont{and}
  \bibinfo{author}{\bibfnamefont{M.~A.} \bibnamefont{Subramanian}},
  \bibinfo{journal}{Phys. Rev. B} \textbf{\bibinfo{volume}{84}},
  \bibinfo{pages}{100403} (\bibinfo{year}{2011}).

\bibitem[{\citenamefont{Zeng et~al.}(2002)\citenamefont{Zeng, Li, Liu, L., and
  Sun}}]{Zeng02}
\bibinfo{author}{\bibfnamefont{H.}~\bibnamefont{Zeng}},
  \bibinfo{author}{\bibfnamefont{J.}~\bibnamefont{Li}},
  \bibinfo{author}{\bibfnamefont{J.~P.} \bibnamefont{Liu}},
  \bibinfo{author}{\bibfnamefont{W.~Z.} \bibnamefont{L.}}, \bibnamefont{and}
  \bibinfo{author}{\bibfnamefont{S.}~\bibnamefont{Sun}},
  \bibinfo{journal}{Nature (London)} \textbf{\bibinfo{volume}{420}},
  \bibinfo{pages}{395} (\bibinfo{year}{2002}).

\bibitem[{\citenamefont{Balamurugan et~al.}(2014)\citenamefont{Balamurugan,
  Das, Zhang, Skomski, and J.}}]{Balamurugan14}
\bibinfo{author}{\bibfnamefont{B.}~\bibnamefont{Balamurugan}},
  \bibinfo{author}{\bibfnamefont{B.}~\bibnamefont{Das}},
  \bibinfo{author}{\bibfnamefont{W.~Y.} \bibnamefont{Zhang}},
  \bibinfo{author}{\bibfnamefont{R.}~\bibnamefont{Skomski}}, \bibnamefont{and}
  \bibinfo{author}{\bibfnamefont{S.~D.} \bibnamefont{J.}}, \bibinfo{journal}{J.
  Phys. Condens. Matter} \textbf{\bibinfo{volume}{26}}, \bibinfo{pages}{064204}
  (\bibinfo{year}{2014}).

\end{thebibliography}

\end{document}